\documentclass[12pt, twoside]{report}

\newcommand{\ThesisTitle}{Scattering Observables from Few-Body Densities and Application in Light Nuclei}
\newcommand{\ThesisAuthor}{Alexander Phillips Long}

\newcommand{\ThesisUniversity}{The George Washington University}
\newcommand{\ThesisCollege}{Columbian College of Arts and Sciences}

\newcommand{\ThesisDegreeType}{Doctor}
\newcommand{\ThesisDegreeDiscipline}{Philosophy}
\newcommand{\ThesisDegree}{\ThesisDegreeType{} of \ThesisDegreeDiscipline}

\newcommand{\ThesisDefenseDate}{June 4, 2026}  
\newcommand{\ThesisGradDate}{August 31, 2026}  
\newcommand{\ThesisCopyrightYear}{2026}

\newcommand{\ThesisPrevDegreeOne}{Bachelor of Science in Physics, May 2019, University of St.\ Thomas}
\newcommand{\ThesisPrevDegreeTwo}{Bachelor of Arts in Mathematics, May 2019, University of St.\ Thomas}

\newcommand{\AdvisorName}{Harald W. Grie{\ss}hammer}
\newcommand{\AdvisorTitle}{Professor of Physics and Director of Graduate Studies}
\newcommand{\AdvisorRole}{Dissertation Director}
\newcommand{\AdvisorCommitteeLine}{\AdvisorName, \AdvisorTitle, \AdvisorRole}

\newcommand{\CommitteeMemberOneName}{Michael Doering}
\newcommand{\CommitteeMemberOneTitle}{Associate Professor of Physics; Deputy Chair}
\newcommand{\CommitteeMemberOneRole}{Committee Member}
\newcommand{\CommitteeMemberOneLine}{\CommitteeMemberOneName, \CommitteeMemberOneTitle, \CommitteeMemberOneRole}

\newcommand{\CommitteeMemberTwoName}{Evangeline J. Downie}
\newcommand{\CommitteeMemberTwoTitle}{Professor of Physics}
\newcommand{\CommitteeMemberTwoRole}{Committee Member}
\newcommand{\CommitteeMemberTwoLine}{\CommitteeMemberTwoName, \CommitteeMemberTwoTitle, \CommitteeMemberTwoRole}

\usepackage[T1]{fontenc}

\usepackage{amsmath,amssymb}
\usepackage{amsthm}

\usepackage{mathtools}
\usepackage{mathrsfs}
\usepackage{bbm}
\usepackage{slashed}
\usepackage{graphicx}
\usepackage{tikz}
\usetikzlibrary{arrows, decorations.markings, decorations.pathmorphing, calc}

\usepackage{xspace}
\usepackage{xstring}
\usepackage{etoolbox}
\usepackage{bigints}
\usepackage{enumitem}
\usepackage{wrapfig}
\usepackage{float}
\usepackage{array}
\usepackage{booktabs}
\usepackage{multirow}
\usepackage{subcaption}
\usepackage{tabularx}
\usepackage[table]{xcolor}
\usepackage{adjustbox}
\usepackage{multicol}
\usepackage{caption}
\usepackage{changepage}
\usepackage{threeparttable}
\usepackage{mathalfa}
\DeclareMathAlphabet{\mathscreuler}{U}{eus}{m}{n}

\usepackage{setspace}
\usepackage{indentfirst}
\usepackage[letterpaper,top=1.in, bottom=2.0cm, inner=2.5cm, outer=2.5cm]{geometry}

\usepackage{fancyhdr}
\usepackage[final,breaklinks=true,
            colorlinks=true,
            pdfborder={0 0 1},
            citecolor={blue},linkcolor={blue},urlcolor={red},
            citebordercolor={1 0 0},linkbordercolor={0 0 1},urlbordercolor={1 0 0},
            pdfpagemode=UseOutlines,
            bookmarks=true,bookmarksopenlevel=4
]{hyperref}
\usepackage{xurl}
\usepackage{tocloft}
\makeatletter
\patchcmd{\tableofcontents}
  {\chapter*}
  {\clearpage \vspace*{0pt} \chapter*}
  {}{}
\makeatother

\usepackage{titlesec}
\titleformat{\chapter}[display]
  {\normalfont\huge\bfseries\raggedright}{\MakeUppercase{\chaptertitlename}\ \thechapter}{20pt}{\Huge}
\titlespacing*{\chapter}{0pt}{50pt}{20pt}

\numberwithin{equation}{chapter}
\numberwithin{figure}{chapter}
\numberwithin{table}{chapter}

\usepackage[
  biblabel  = brackets,
  style     = phys,
  citestyle = numeric-comp,
  sorting   = none,
  backend   = biber,
  natbib    = true
]{biblatex}
\AtEveryBibitem{%
  \clearfield{note}%
  \clearlist{note}%
  \clearfield{language}%
  \clearlist{language}%
  \clearfield{file}%
  \clearlist{file}%
  \iffieldundef{date}{}{%
    \clearfield{year}%
    \clearfield{month}%
    \clearfield{day}%
  }%
  \clearfield{urldate}%
  \clearlist{urldate}%
  \clearfield{urlseen}%
  \clearlist{urlseen}%
}

\DefineBibliographyStrings{english}{urlseen = {accessed}}

\newcommand{\cEFT}{$\chi$EFT\xspace}
\newcommand{\Ep}{E_\pi}

\newcommand{\mpi}{m_\pi}

\newcommand{\mz}{m_\pi}

\newcommand{\wth}{\omega_{th}}

\DeclarePairedDelimiterX\braket[2]{\langle}{\rangle}{#1\,\delimsize\vert\,\mathopen{}#2}
\DeclarePairedDelimiterX\brkt[2]{\langle}{\rangle}{#1\,\delimsize\vert\,\mathopen{}#2}

\newcommand\ddfrac[2]{\frac{\displaystyle #1}{\displaystyle #2}}
\newcommand{\dd}{\mathrm{d}}

\newcommand{\etal}{\textit{et al.}~}

\newcommand{\HeT}{{}^{3} \mathrm{He}\xspace}

\newcommand{\MeV}{\mathrm{MeV}}

\newcommand{\fmin}{\mathrm{fm}^{-1}}

\newcommand{\chiEFT}{$\chi$EFT\xspace}

\newcommand{\p}{\partial}

\newcommand{\CG}[6]{\langle {#1} {#2}%
  {\IfSubStr{#4}{+}{(#4)}{\IfSubStr{#4}{-}{(#4)}{#4}}}
  {\IfSubStr{#5}{+}{(#5)}{\IfSubStr{#5}{-}{(#5)}{#5}}}  |
   {#3} {\IfSubStr{#6}{+}{(#6)}{\IfSubStr{#6}{-}{(#6)}{#6}}} \rangle}

\newcommand{\kv}{\vec{k}}

\newcommand{\kvp}{\vec{k}^\prime}

\newcommand{\pv}{\vec{p}}

\newcommand{\vv}{\vec{v}}

\newcommand{\uot}{_{12}}
\newcommand{\ez}{\hat{\mathrm{e}}_z}
\newcommand{\ey}{\hat{\mathrm{e}}_y}

\newcommand{\meqref}[1]{eq.~\eqref{#1}}

\newcommand{\Mnucl}{M_{\mathrm{nucl}}}

\def\backmatter{%
    \setcounter{section}{0}%
    \renewcommand{\thesection}{\Alph{section}}%
}

\title{\ThesisTitle}
\author{\ThesisAuthor}

\def\mydirs{intro,theory,QCD,TDA,SRG,usage,compton,piphoto,pion,outlook}

\def\mygraphicspath{}
\renewcommand*{\do}[1]{\edef\mygraphicspath{\mygraphicspath{#1/}}}
\expandafter\forcsvlist\expandafter\do\expandafter{\mydirs}
\edef\mygraphicspath{\mygraphicspath{appendix/}}
\expandafter\graphicspath\expandafter{\mygraphicspath}

\begin{document}

\pagenumbering{roman}

\begin{titlepage}
\thispagestyle{empty}
\begin{singlespace}
\begin{center}
    \vspace*{0.75in}
    \ThesisTitle

    \vspace{3\baselineskip}

    by \ThesisAuthor

    \vspace{3\baselineskip}

    \ThesisPrevDegreeOne\\
    \ThesisPrevDegreeTwo

    \vspace{2\baselineskip}

    A Dissertation submitted to\\

    \vspace{3\baselineskip}

    The Faculty of\\
    \ThesisCollege\\
    of \ThesisUniversity\\
    in partial fulfillment of the requirements\\
    for the degree of \ThesisDegree

    \vspace{3\baselineskip}

    \ThesisGradDate

    \vspace{3\baselineskip}

    Dissertation directed by

    \vspace{1\baselineskip}

    \AdvisorName\\
    \AdvisorTitle
\end{center}
\end{singlespace}
\end{titlepage}

\clearpage
\setcounter{page}{2}

\noindent The \ThesisCollege{} of \ThesisUniversity{} certifies that \ThesisAuthor{} has passed the Final Examination of the degree of \ThesisDegree{} as of \ThesisDefenseDate. This is the final and approved form of the dissertation.

\begin{center}
{\setstretch{1}\vspace{1.5\baselineskip}}

\ThesisTitle

{\setstretch{1}\vspace{1.5\baselineskip}}

\ThesisAuthor
\end{center}

{\setstretch{1}\vspace{1.5\baselineskip}}

\begin{singlespace}

\noindent Dissertation Research Committee:

{\setstretch{1}\vspace{0.5\baselineskip}}

\begin{adjustwidth}{0.5in}{0in}
    \AdvisorCommitteeLine
\end{adjustwidth}

{\setstretch{1}\vspace{0.5\baselineskip}}

\begin{adjustwidth}{0.5in}{0in}
    \CommitteeMemberOneLine
\end{adjustwidth}

{\setstretch{1}\vspace{0.5\baselineskip}}

\begin{adjustwidth}{0.5in}{0in}
    \CommitteeMemberTwoLine
\end{adjustwidth}
\end{singlespace}

\clearpage

\begin{singlespace}
\begin{center}
\vspace*{\fill}

\copyright\ Copyright \ThesisCopyrightYear{} by \ThesisAuthor\\
All rights reserved

\vspace*{\fill}
\end{center}
\end{singlespace}

\clearpage

\addcontentsline{toc}{section}{Dedication}
\begin{center}
\textbf{Dedication}

\vspace{0.5\baselineskip}
\end{center}

\vspace*{\fill}
\begin{center}
Dedicated to Willis F. Long PhD\\
Thank you for your presence, inspiration and patience.
\end{center}
\vspace*{\fill}

\clearpage

\addcontentsline{toc}{section}{Abstract of Dissertation}
\begin{singlespace}
\begin{center}
\textbf{Abstract of Dissertation}

{\setstretch{1}\vspace{2\baselineskip}}

\ThesisTitle
\end{center}

\vspace{-0.5\baselineskip}
\end{singlespace}

The Transition Density Amplitude (TDA) method of
Grie{\ss}hammer \textit{et al.}\ is applied to calculate
scattering observables for Compton scattering,
neutral pion photoproduction, and
elastic pion scattering off the light nuclei ${}^3$H,
${}^3$He, ${}^4$He, and ${}^6$Li. In this approach, the
scattering amplitude factorizes into an irreducible
few-body kernel, encoding the interaction of the probe
with the active nucleons, and a transition density
amplitude that encapsulates the nuclear structure
information of the spectator nucleons. The TDAs are
computed once for a given nucleus and a given momentum transfer, and subsequently
combined with the appropriate kernel for any elastic
reaction, thereby providing a unified and modular
framework for treating multiple processes within Chiral
Effective Field Theory ($\chi$EFT).
All calculations are carried out within the publicly available Fortran code
suite \texttt{DensityScattering}~\cite{long_2026},
developed as part of the present work; a researcher
wishing to compute a new reaction need only supply the
process-specific kernel, while all infrastructure for
density handling, integration, quantum-number summation,
and output is provided by the existing framework.

Nuclear wave functions are obtained from the semilocal
momentum-space regularized chiral potential ($\chi$SMS) of
Reinert, Krebs, and Epelbaum at cutoffs
$\Lambda = 450$~MeV and $500$~MeV, with Similarity
Renormalization Group (SRG) transformations, and back-transformation to physical momenta employed to
extend the calculations to $A = 6$. The SRG-induced
uncertainties are validated against exact results for
${}^4$He, where deviations are found to be at the $2\%$
level or below. Cutoff selection is informed by the
Bayesian convergence analysis of Millican \textit{et al.}~\cite{SMS400550},
which identifies $\Lambda = 450$ and $500$~MeV as the
values exhibiting statistically consistent convergence
patterns.

For Compton scattering on ${}^6$Li, differential cross
sections are computed at photon energies of $60$, $75$,
$86$, and $100$~MeV using one- and two-body kernels through
$\mathcal{O}(e^2\delta^3)$ in $\Delta(1232)$-full
$\chi$EFT, and are compared with available data. 
The results presented have uncertainties on the $6\%$ level and are in tension with experiment.

For threshold neutral pion photoproduction, both one-body
and two-body contributions to the $S$-wave multipole
amplitudes $E_{0+}$ and $L_{0+}$ are calculated, including
$\mathcal{O}(q^4)$ static corrections to the two-body
kernel. The results for the trinucleon systems are in good
agreement with the calculations of Lenkewitz
\textit{et al.}~\cite{Lenke2013, Lenke2011, LenkeThesis}.
For ${}^6$Li, the threshold amplitude agrees with the
Saskatchewan Accelerator Laboratory measurement of Bergstrom
\textit{et al.}~\cite{Bergstrom1999} at the $1.8\sigma$ level,
reproducing the same fractional offset between \chiEFT and
data found for the deuteron~\cite{Beane_1997, Bergstrom1998},
where the chiral prediction provided an early landmark
confirmation of the two-body pion-production mechanism. A
systematic factor of $\sqrt{2}$ discrepancy with the results
of Braun is also identified for ${}^6$Li and traced to an
error in the normalization of the spin-$1$ angular momentum
matrices used to extract the form factors.
A variable substitution technique is developed
to treat the removable moving singularity of the form
$1/\vec{q}\,^2$ that arises in the two-body propagator;
this method maps the singularity to the origin of the
integration variable, where it is canceled by the Jacobian
of the spherical measure, replacing the regulator
extrapolation employed in prior
calculations~\cite{Lenke2013,BraunThesis} and eliminating
the associated systematic uncertainty.
Across all three nuclei, the two-body contribution
dominates the total production amplitude, consistent with
the suppression of the leading one-body contact term for
neutral pion channels. Above threshold, the one-body
contribution to pion photoproduction is computed at a
photon energy of $170$~MeV using SAID partial-wave
amplitudes as input, demonstrating
the capability of the TDA framework to produce predictions
across a range of energies.

Elastic pion-nucleus scattering lengths are computed for
${}^3$He, ${}^4$He, and ${}^6$Li using one-body kernels
derived from SAID partial-wave amplitudes and two-body
kernels constructed from the leading diagrams of Liebig
\textit{et al.}~\cite{Liebig_2011}, including the numerically enhanced
triple-scattering contribution. The two-body sector
dominates the total scattering lengths for all nuclei
considered, and an isospin decomposition of the results
confirms that isospin remains a good approximate symmetry
at the level of accuracy attained. Above threshold, the
one-body elastic pion scattering amplitude is evaluated at
a center-of-mass pion energy of $170$~MeV using SAID WI08
partial-wave amplitudes, exhibiting the angular structure
produced by the $P_{33}$ partial wave near the
$\Delta(1232)$ resonance.

\clearpage

\thispagestyle{plain}

\vspace*{-1.5\baselineskip}
\begin{center}
    \textbf{Table of Contents}
\end{center}

\vspace{-3\baselineskip}

\begingroup
\renewcommand{\contentsname}{}
\tableofcontents
\endgroup

\clearpage
\phantomsection
\addcontentsline{toc}{chapter}{List of Figures}

\thispagestyle{plain}

\vspace*{-1.5\baselineskip}
\begin{center}
    \textbf{List of Figures}
\end{center}

\vspace{-3\baselineskip}

\begingroup
\renewcommand{\listfigurename}{}
\listoffigures
\endgroup

\clearpage
\phantomsection
\addcontentsline{toc}{chapter}{List of Tables}

\thispagestyle{plain}

\vspace*{-1.5\baselineskip}
\begin{center}
    \textbf{List of Tables}
\end{center}

\vspace{-3\baselineskip}

\begingroup
\renewcommand{\listtablename}{}
\listoftables
\endgroup

\clearpage
\pagenumbering{arabic}
\setcounter{page}{1}

\let\oldsubsubsection\subsubsection
\let\oldsubsection\subsection
\let\oldsection\section
\let\section\chapter
\let\subsection\oldsection
\let\subsubsection\oldsubsection

\renewcommand*{\do}[1]{\include{#1/#1}}
\expandafter\forcsvlist\expandafter\do\expandafter{\mydirs}

\let\section\oldsection
\let\subsection\oldsubsection
\let\subsubsection\oldsubsubsection

\clearpage
\phantomsection
\addcontentsline{toc}{chapter}{Bibliography}

\begingroup
  \sloppy
  \setlength{\emergencystretch}{3em}
  \Urlmuskip=0mu plus 1mu\relax
  \printbibliography
\endgroup

\clearpage
\appendix
\phantomsection
\addcontentsline{toc}{chapter}{Appendices}
\chapter*{Appendices}
\markboth{Appendices}{Appendices}

\makeatletter
\@addtoreset{equation}{section}
\@addtoreset{figure}{section}
\@addtoreset{table}{section}
\makeatother
\renewcommand{\theequation}{\thesection.\arabic{equation}}
\renewcommand{\thefigure}{\thesection.\arabic{figure}}
\renewcommand{\thetable}{\thesection.\arabic{table}}

\backmatter
\section{Conventions}
\label{sec:conventions}
Unless explicitly stated otherwise, the conventions
in this work follow those of Bernard \etal
\cite{BERNARD_1995}. Natural units $\hbar = c = 1$
are employed throughout.
\begin{equation}
  \begin{split}
  g_A &= 1.26\;, \\
  f_\pi &= 92.42~\MeV\;,
    \quad\text{(pion decay constant)} \\ 
  \gamma_5 &= \gamma^5
    = i\,\gamma^0\gamma^1\gamma^2\gamma^3\;, \\
  g^{\mu \nu} &= \mathrm{Diag}(+,-,-,-).
  \end{split}
  \label{eq:conventions}
\end{equation}
The gamma matrices $\gamma^\mu$ are $4\times 4$
matrices which can be defined by the necessary and
sufficient condition
\begin{equation}
  \left\{\gamma^\mu, \gamma^\nu\right\}=\gamma^\mu \gamma^\nu+\gamma^\nu \gamma^\mu=2 g^{\mu \nu} \mathbbm{1}.
\end{equation}
The Gell-Mann matrices are given by
\begin{equation}
\begin{aligned}
& \lambda_1=\left(\begin{array}{lll}
0 & 1 & 0 \\
1 & 0 & 0 \\
0 & 0 & 0
\end{array}\right) \quad \lambda_2=\left(\begin{array}{ccc}
0 & -i & 0 \\
i & 0 & 0 \\
0 & 0 & 0
\end{array}\right) \quad \lambda_3=\left(\begin{array}{ccc}
1 & 0 & 0 \\
0 & -1 & 0 \\
0 & 0 & 0
\end{array}\right) \\
& \lambda_4=\left(\begin{array}{lll}
0 & 0 & 1 \\
0 & 0 & 0 \\
1 & 0 & 0
\end{array}\right) \quad \lambda_5=\left(\begin{array}{ccc}
0 & 0 & -i \\
0 & 0 & 0 \\
i & 0 & 0
\end{array}\right) \\
& \lambda_6=\left(\begin{array}{lll}
0 & 0 & 0 \\
0 & 0 & 1 \\
0 & 1 & 0
\end{array}\right) \quad \lambda_7=\left(\begin{array}{ccc}
0 & 0 & 0 \\
0 & 0 & -i \\
0 & i & 0
\end{array}\right) \quad \lambda_8=\frac{1}{\sqrt{3}}\left(\begin{array}{ccc}
1 & 0 & 0 \\
0 & 1 & 0 \\
0 & 0 & -2
\end{array}\right) .
\end{aligned}
\label{eq:GellMann}
\end{equation}
The Pauli matrices are given by
\begin{equation}
\sigma_x=\begin{pmatrix}
0 & 1 \\
1 & 0
\end{pmatrix} \quad
\sigma_y=\begin{pmatrix}
0 & -i \\
i & 0
\end{pmatrix} \quad
\sigma_z=\begin{pmatrix}
1 & 0 \\
0 & -1
\end{pmatrix}\;.
\label{eq:Pauli}
\end{equation}
The spin-1 matrices are given by
\begin{equation}
S_x=\frac{1}{\sqrt{2}}\begin{pmatrix}
0 & 1 & 0 \\
1 & 0 & 1 \\
0 & 1 & 0
\end{pmatrix} \quad
S_y=\frac{1}{\sqrt{2}}\begin{pmatrix}
0 & -i & 0 \\
i & 0 & -i \\
0 & i & 0
\end{pmatrix} \quad
S_z=\begin{pmatrix}
1 & 0 & 0 \\
0 & 0 & 0 \\
0 & 0 & -1
\end{pmatrix}\;.
\label{eq:Spin1}
\end{equation}
Throughout this work, uncertainty propagation is
performed under the assumption of uncorrelated
variables. For a function $z = f(x,y)$ with
associated uncertainties $\sigma_x$ and $\sigma_y$,
the uncertainty on $z$ is estimated via
\begin{equation}
  \sigma_z = \sqrt{
  \left( \frac{\p z}{\p x}   \right)^2 \sigma_x^2 +
  \left( \frac{\p z}{\p y}   \right)^2 \sigma_y^2
}\;.\label{eq:propagation}
\end{equation}

\section{Converting Between Densities}
\label{sec:convert}
The Mandelstam variables $t$ and $s$ define the
density required for the calculation; the values of
$t$ and $s$ are then converted to the equivalent
center of mass energy and angle for Compton
scattering, and for historical reasons these Compton
energies and angles are the quantities recorded in
the database available through the
\href{https://pypi.org/project/nucdens/}{nucdens}
Python package.
\subsection{Pion-Photoproduction Density Conversion}
\label{sec:piphoto_convert}
Let the subscript $\gamma\gamma$ denote Compton
scattering, and let the subscript $\mu$ denote some
other reaction; later $\mu=\gamma\pi$ denotes
pion photoproduction, and $\mu=\pi\pi$ denotes
pion scattering.
In Compton scattering, let the angle of the outgoing
Compton photon be $\phi$ and the energy of the
incoming photon be $\omega$.
Now consider the Mandelstam variables as functions of
the energy $E$ and the angle $\theta$ in the center
of mass frame.
To transform between reactions, the Mandelstam
variables are equated (using function notation)
\begin{equation}
  s_{\gamma\gamma}(\omega,\phi)=s_\mu(E_\mu,\theta_\mu),\qquad
  t_{\gamma\gamma}(\omega,\phi)=t_\mu(E_\mu,\theta_\mu)\;,
\end{equation}
and then inverted to find expressions for
\begin{align}
  \omega(E_\mu,\theta_\mu)\label{omegaExp}\\
  \phi(E_\mu,\theta_\mu)\;.\label{phiExp}
\end{align}
Expressing $s_\mu$ and $s_{\gamma\gamma}$ in terms
of incoming momenta eliminates the $\phi$ and
$\theta_\mu$ dependence of $s$.
The Compton scattering Mandelstam variable $t$ is
given by
\begin{equation}
\begin{split}
  k&=(\omega,\omega\ez)\\
  k^\prime&=\left(\omega, \omega[\sin{\phi}\,\ey+\cos{\phi}\,\ez] \right)\\
  \implies t_{\gamma\gamma}&=(k-k^\prime)^2=-2\omega^2 (1-\cos{\phi})\;.
\end{split}
\end{equation}
As an example, consider pion photoproduction; the
particles present in the incoming state are exactly
the same as in the corresponding Compton reaction
with the same incident photon energy $\omega$.
The pion-photoproduction Mandelstam variable $t$ is
then given by
\begin{equation}
  \begin{split}
  k&=(\omega, \omega \ez)\\
  k^\prime&=\left(\Ep, |\kvp|(\sin{\theta}\,\ey+\cos{\theta}\,\ez)\right)\\
  \implies t_{\gamma\pi}&=(k-k^\prime)^2 =-|\kvp|^2 + E_\pi^{2}-2\omega_{\gamma\pi} E_\pi+ 2 |\kvp|\, \omega_{\gamma\pi} \cos{\theta}\;,
  \end{split}
\end{equation}
where $\Mnucl$ denotes the nucleus mass, $\mpi$
the pion mass, $E_\pi$ is the energy of the outgoing
pion, and $\kvp$ is the outgoing pion momentum.
To convert the energy, note that
\begin{align}
  s_{\gamma\gamma}=(k+p)^2&= \left[ \left(\sqrt{\Mnucl^2+\omega^2}, -\omega \ez\right) + (\omega,\omega \ez ) \right]^2\\
                          &= \left( \omega +\sqrt{\Mnucl^2+\omega^2} \right)^2=s_{\gamma \pi}\\
  \implies \omega_{\gamma\gamma}&=\omega_{\gamma\pi}\equiv\omega\;. \label{omegaExpSol}
\end{align}
Thus \meqref{omegaExpSol} provides the (trivial)
expression for \meqref{omegaExp}, and the indices
$\gamma\pi$ and $\gamma\gamma$ on $\omega$ are
henceforth dropped because they are redundant.
Solving for $\phi$,
\begin{align}
  t_{\gamma\pi}& =
  -|\kv^\prime|^2 + E_\pi^{2}-2\omega E_\pi+ 2 |\vec{k}^\prime|\, \omega \cos{\theta_\pi}
  =-2\omega^2 (1-\cos{\phi})=t_{\gamma\gamma}\\
  \implies & \phi=\arccos \frac{-2 E_\pi \omega +2 \omega  \cos \theta_\pi  \sqrt{E_\pi^2-m_\pi^2}+m_\pi^2+2 \omega ^2}{2 \omega ^2}\;.\label{phiExpSol}
\end{align}
Thus \meqref{phiExpSol} is the solution to
\meqref{phiExp}, and one can use
\begin{equation}
  \omega =\omega_{\gamma\pi}= \ddfrac{s - \Mnucl^2}{2 \sqrt{s}}, \quad E_\pi = \ddfrac{s+ m_\pi^2 - \Mnucl^2}{2\sqrt{s}}\;.
\end{equation}
One interesting point is at threshold:
\begin{align}
	s=(k+p)^2 & = \left[ \left(\sqrt{\Mnucl^2+\wth^2},0,0,-\wth\right) + (\wth, 0,0,\wth) \right]^2 \nonumber\\
	          & = \left[ (\Mnucl,0,0,0) + (\mz,0,0,0) \right] ^2=(k'+p')^2                                    \\
	\implies  & (\sqrt{\Mnucl^2 +\wth^2}+\omega, 0,0, 0)^2 =(\Mnucl+ \mz)^2 \nonumber\\
	\implies  & \wth = \frac{\mz (\mz + 2\Mnucl)}{2 (\mz +\Mnucl)}\;.
\end{align}
For example, for a $\HeT$ target with
$E_\pi=m_{\pi^0}=134.987\, \MeV$ and
$\Mnucl=2808.4\,\MeV$, the result is
$\phi=59.98^\circ$ and
$\omega=131.857\,\MeV$.
\subsection{Pion Scattering Density Conversion}
\label{sec:pion_convert}
For pion scattering, a procedure analogous to that
of pion photoproduction is carried out, converting
between the kinematics of Compton scattering and
pion scattering in order to select the correct
density from the
\href{https://pypi.org/project/nucdens/}{nucdens}
database. The fundamental dependence is on the
Mandelstam variables $s$ and $t$, so given a pion
energy $E_\pi$ and outgoing pion angle $\theta_\pi$,
the Compton Mandelstam variables
$s_{\gamma\gamma}$ and $t_{\gamma\gamma}$ are sought
as functions of Compton energy $\omega$ and outgoing
Compton angle $\phi$ such that
\begin{equation}
  s_{\gamma\gamma}(\omega,\phi)=s_{\pi}(E_{\pi},\theta_{\pi}),\qquad
  t_{\gamma\gamma}(\omega,\phi)=t_{\pi}(E_{\pi},\theta_{\pi})\;,
\end{equation}
As before, the goal is to invert these relations
and obtain expressions for $\omega$ and $\phi$.
Letting $k$ denote the pion four-momentum,
$\Mnucl$ the nucleus mass, and $\mpi$ the pion
mass, the $s$-channel relation reads
\begin{align}
  s_\pi&= (p+k)^2= \left( \sqrt{\mpi^2 +\kv^2} + \sqrt{\Mnucl^2+ \kv^2}\right)^2
  = \left( \omega+ \sqrt{\Mnucl^2 +\omega^2} \right)^2=s_{\gamma\gamma}\;.
\end{align}
Solving yields
\begin{equation}
  \omega= \frac{1}{2 \left( \Mnucl^2 -\mpi^2 \right) }
  \left[ 2 \Mnucl^2 \sqrt{\kv^2+\mpi^2} -\mpi^2 \left(\sqrt{\kv^2+\mpi^2}+ \sqrt{\kv^2+\Mnucl^2}\right) \right]\;.\label{piomegaExp}
\end{equation}
For the angle, equating $t_\pi = t_{\gamma\gamma}$
gives
\begin{equation}
  t_{\pi}= -2 \kv^2 (1- \cos{\theta_\pi})= -2 \omega^2 (1-\cos{\phi})\;.\label{pit}
\end{equation}
Solving yields
\begin{equation}
  \phi = \arccos \left[ 1+ \frac{\kv^2}{\omega^2} \left( \cos{\theta_\pi}-1 \right) \right]\;,\label{piPhi}
\end{equation}
where $E_\pi^2-m_\pi^2=\kv^2$.
At threshold, $\kv=0$, and \meqref{piPhi}
immediately yields $\phi=0$.

\section{Integration Techniques}\label{sec:Inte}

Evaluating tree-level diagrams in the TDA formalism
requires the integration over incoming and outgoing
momenta.
\begin{equation}
  \int \dd^3 p\, \dd^3 v f(\pv,\vv)= \int \dd r_p \dd r_v \dd \Omega_p \dd \Omega_v f(r_p,r_v,\Omega_p, \Omega_v)\;.
\end{equation}
This requires an angular and a radial integration.

\subsection{Angular Integration}
The angular integrations are done with the
Lebedev-Laikov scheme \cite{Lebe}, which, given a
degree $N$ selects weights $W_i$ and points
$\theta_i, \varphi_i$ such that
\begin{align}
  \int \dd \Omega f(\Omega)&= \int_0^\pi \sin{\theta} \,\dd \theta \int_0^{2\pi} f(\theta, \varphi) \dd \varphi
                           \approx 4\pi \sum_{i=1}^N W_i f(\theta_i,\varphi_i)\;.
\end{align}
The grid points and weights in this scheme are
constructed such that they are invariant under the
octahedral rotation group with inversion.
The advantage of this construction is the absence
of a preferred orientation, which would otherwise
impair convergence due to the $\sin{\theta}$
weighting in the integrand.
Notably, in this scheme, only particular values of
$N$ are allowed, which correspond to certain
octahedral group symmetries.
For many $N$ the points $\theta_i, \varphi_i$
correspond to vertices of regular polyhedra, but the
construction of the points and weights order by order
varies by $N$ and is therefore beyond the scope of
this discussion.
For a simple example, at $N=6$ points the grid values
are $(x,y,z)=(\pm1,0,0),\;(0,\pm1,0), (0,0,\pm1)$,
and $W_i=\frac{1}{6} $ for all $i$.
A minimal implementation of the Lebedev-Laikov scheme
can be found in the Python
``\href{https://docs.scipy.org/doc/scipy/reference/generated/scipy.integrate.lebedev_rule.html}{SciPy}''
package, and it is implemented in the
\texttt{DensityScattering} code \cite{long_2026}
subroutines.

\subsection{Radial Integration}
Radial integration is done with the Gauss-Legendre
scheme which approximates an integral through the
summation
\begin{equation}
  \int_{-1}^{1} f(x) \dd x \approx \sum_{i=1}^n w_i f(x_i)\;,
\end{equation}
where $x_i$ is the $i$-th root of the Legendre
polynomial of degree $n$ and the weights $w_i$ are
\begin{equation}
  w_i= \frac{2}{(1-x_i^2) \left[ P'_n(x_i) \right]^2}\;.
\end{equation}
In the present TDA implementation, the radial
integration is partitioned into three regions,
with the Gauss-Legendre points $x_i$ being mapped to
these regions.
The number of integration points $n_1$ and $n_2$ are
specified by the user in an input file,
and a total of three different Gauss Legendre
integrations are done.
These integrations cover the regions
$X \in [0, P_1]$ with $n_1/2$ points,
$X \in [P_1, P_2]$ with $n_1/2$ points, and
$X \in [P_2, P_3]$ with $n_2$ points.
$P_3$ acts as the upper bound cutoff of the radial
integral.
For the region $X\in [0,P_2]$
\begin{align}
  X_i = \ddfrac{ (1+x_i)}{P_1-\left(\frac{1}{P_1}-\frac{2}{P_2}\right)x_i}\;,
\end{align}
where $x_i$ come from the Gauss-Legendre scheme on
the $[-1,1]$ interval. This ensures equal points are
generated in the region $[0,P_1]$ and $[P_1, P_2]$.
For the region $X\in [P_2,P_3]$
\begin{equation}
X_i = \frac{P_3 + P_2}{2}
  + \frac{P_3 - P_2}{2}\, x_i\;.
\end{equation}
The corresponding Jacobian is also applied to the
weights.
For $P_i$ typical values are
$P_1= 1.1\,\fmin$,
$P_2= 5.0\,\fmin$, and
$P_3= 15.0\,\fmin$.
\subsection{Integration and Angular Momentum Convergence}
\label{sec:InteConverge}
The calculations presented in this work for radial,
Gauss-Legendre integration utilize $n_1=14$, and for
the second region $n_2=2$. Angular integration is
done with a Lebedev-Laikov grid of order $N=50$.
In addition, the convolution with the TDA involves a
sum over $j\uot = 0, 1, \ldots$ which must be
truncated. From convergence analysis,
$j_{12\,\text{max}} = 1$ has been selected as the
maximum value of $j\uot$.
\begin{table}[H]
\centering
\begin{tabular}{l|c|c}
    &$\left(F_T^{(a)} - F_T^{(b)}\right)/2$& $\left(F_L^{(a)} - F_L^{(b)}\right)/2$\\
  \hline$n_1=10,n_2=2, N=30, j_{12\,\text{max}}=1$& $-14.94112$ & $-12.20165$ \\
  \hline$n_1=14,n_2=2, N=50, j_{12\,\text{max}}=1$& $ -15.13542$& $- 12.22171$\\
  \hline$n_1=14,n_2=2, N=50, j_{12\,\text{max}}=2$& $-15.08532$& $- 12.17784$\\
  \hline$n_1=16,n_2=4, N=72, j_{12\,\text{max}}=1$& $-15.16041 $& $- 12.20165$\\
\end{tabular}
\caption[Pion photoproduction ${}^3$He form factor convergence]{
  Pion Photoproduction $\HeT$ form factors for various numerical settings with $\chi$SMS potential with a cutoff $\Lambda=450 \MeV$.
  The values of the second row, $n_1=14,\,n_2=2,\,N=50, j_{12\,\text{max}}=1$ are used throughout this work.
}
\label{HeTConvTable}
\end{table}
This selection of $n_1, n_2$ and $N$ is justified by
the numerical convergence presented in
Table~\ref{HeTConvTable}, where pion
photoproduction form factors are computed for
various parameter settings.
Increasing the integration grid to
$n_1=16, n_2=4, N=72$ changes the result by
$0.16\%$. Similarly, using the default integration
grid, going from $j_{12\,\text{max}}=1$ to
$j_{12\,\text{max}}=2$ results in a change of
$0.33\%$. These numerical uncertainties are
significantly smaller than those arising from the
truncation of the \cEFT expansion.

\end{document}